\documentclass[superscriptaddress,twocolumn,english,floatfix,aps,pra,amsmath,amssymb,longbibliography]{revtex4-2}
\usepackage[T1]{fontenc}
\usepackage[utf8]{inputenc}
\usepackage{babel}
\usepackage{comment}
\usepackage{color}
\usepackage{times}
\usepackage{epsfig}
\usepackage{subfigure}
\usepackage{amsmath}
\usepackage{amssymb}
\usepackage{graphicx}
\usepackage[colorlinks=true]{hyperref}
\hypersetup{citecolor=blue,linkcolor=blue,urlcolor=blue}

\begin{document}
	
\title{Sign inversion in the lateral van der Waals force}
	
\author{Edson C. M. Nogueira}
\email{edson.moraes.nogueira@icen.ufpa.br}
\affiliation{Faculdade de F\'{i}sica, Universidade Federal do Par\'{a}, 66075-110, Bel\'{e}m, Par\'{a}, Brazil}
	
\author{Lucas Queiroz}
\email{lucas.silva@icen.ufpa.br}
\affiliation{Faculdade de F\'{i}sica, Universidade Federal do Par\'{a}, 66075-110, Bel\'{e}m, Par\'{a}, Brazil}
	
\author{Danilo T. Alves}
\email{danilo@ufpa.br}
\affiliation{Faculdade de F\'{i}sica, Universidade Federal do Par\'{a}, 66075-110, Bel\'{e}m, Par\'{a}, Brazil}
\affiliation{Centro de F\'{i}sica, Universidade do Minho, P-4710-057, Braga, Portugal}
	
\date{\today}
	
%%%%%%%%%%%%%%%%%%%%%%%%%%%%%%%%%%%%%%%%%%%%%%%%%%%%%%%%%%%%%%%
\begin{abstract}
We consider a single slight protuberance in a perfectly conducting plane, and investigate the van der Waals (vdW) interaction between this surface and a neutral polarizable particle.
When the protuberance is sufficiently smooth, so that the proximity force approximation (PFA) is well applicable, for a fixed distance of the particle from the plane, the lateral vdW force always points to the protuberance.
On the other hand, by making calculations valid beyond the PFA, we show that nontrivial geometric effects arise when we consider an anisotropic particle, and manipulate the ratio between the characteristic widths of the protuberance and the fixed particle-plane distance.
We predict that, as this ratio decreases, a sign inversion in the lateral vdW force can occur, in the sense that, instead of pointing to the protuberance, in certain situations the lateral force points to the opposite direction.
Moreover, we show that even when such a sign inversion in the lateral vdW force does not occur for a single protuberance, it can arise when two or more protuberances are put together, distinguishing between sign inversions originated by individual or collective effects.
In addition, we show that all these effects have their classical counterparts, involving a neutral particle with a permanent electric dipole moment.
The prediction of such geometric effects on the lateral vdW force may be relevant for a better controlling of the interaction between a particle and a corrugated surface in classical and quantum physics.
\end{abstract}

\maketitle
%%%%%%%%%%%%%%%%%%%%%%%%%%%%%%%%%%%%%%%%%%%%%%%%%%%%%%%%%%%%%%%%%%%%%%%%%%%%%%%%%%%%%%%%%%%%

%
\section{Introduction}
\label{sec-introduction}
In 1873, van der Waals proposed an equation of state for real gases,
which included a term related to intermolecular attractive forces \cite{Waals-Thesis-1873}.
For the case of two non-polar molecules, the intermolecular forces,
usually called dispersion van der Waals (vdW) forces,
are due to the quantum fluctuations in the distributions of charges and
currents in these molecules, and thus, have an electromagnetic nature \cite{Eisenschitzand-London-ZeitPhys-1930,London-ZeitPhys-1930}.
The influence of retardation effects, 
related to the speed of light, on these dispersive forces can be derived with the formalism of quantum electrodynamics 
\cite{Casimir-Polder-Nature-1946,Casimir-Polder-PhysRev-1948, Feinberg-Sucher-PRA-1970}.
As examples of dispersive forces, one can cite the attraction between
two perfectly conducting parallel plates \cite{Casimir-ProcKonNedAkadWet-1948},
and between a neutral polarizable particle and a conducting plane surface \cite{Casimir-Polder-Nature-1946,Casimir-Polder-PhysRev-1948,Lennard-Jones-TransFarSoc-1932}, being these forces relevant in physics, chemistry, biology and engineering \cite{Dimopoulos-PRD-2003,Parsegian-Book-2006,Woods-RMP-2016},
with the progress in the precision of the experiments opening possibilities for applications of these forces in micro and nanotechnology \cite{Ball-Nature-2007,Rodriguez-Capasso-PRL-2010,Rodriguez-NaturePhotonics-2011,Keil-JourModOpt-2016,Gong-Nanophotonics-2020,Stange-PhysicsToday-2021}.
In general, dispersive forces depend on the boundary conditions that material bodies impose on the electromagnetic field, by means of their material properties and geometry
\cite{Israelachvili-PRSL-1972,Milonni-QuantumVacuum-1994,Bordag-Book-2009,Israelachvili-Book-2011,Buhmann-DispersionForces-I,Buhmann-DispersionForces-II,Souza-AJP-2013,Passante-Symmetry-2018} (also see Refs. \cite{Klimchitskaya-Universe-2020,Laliotis-AVSQuantumScience-2021,Woods-AppliedSciences-2021} and references therein).
In this context, nontrivial aspects of the dispersive forces have attracted growing interest, 
as, for instance, those related to the control of the sign of the force,
from an attractive to a repulsive behavior \cite{Dzyaloshinskii-Lifshitz-Pitaevskii-SPU-1961, Boyer-PhysRev-1968,Feinberg-Sucher-PRA-1970,Boyer-PRA-1974,Kupiszewska-JMP-1993, Farina-Santos-Tort-JPA-2002,Farina-Santos-Tort-AJP-2002,Leonhardt-NJP-2007, Munday-Nature-2009,Levin-PRL-2010,Eberlein-PRA-2011,Milton-JPA-2012,Shajesh-PRA-2012, Hu-JN-2016, Buhmann-IJMPA-2016, Abrantes-PRA-2018,Sinha-PRA-2018,Venkataram-PRA-2020,Maghrebi-PRD-2011, Kenneth-PRL-2002, Adhikari-PRA-2017, Hoye-PRA-2018},
existing several situations where the presence of an anisotropic particle has a fundamental role in such a sign inversion \cite{Levin-PRL-2010,Eberlein-PRA-2011,Marchetta-PRA-2021,Buhmann-IJMPA-2016, Abrantes-PRA-2018}.

When corrugations are considered in Casimir interactions, lateral dispersive forces appear and nontrivial geometric effects can be predicted, whose understanding can be important, for instance, to achieve a higher degree of control of the interaction between nanogratings (see, for instance, Refs. \cite{Tang-NaturePhotonics-2017, Wang-NatureComm-2021}), to possibly create micro/nanodevices which operate through the lateral force (see, for instance, Refs. \cite{Ashourvan-PRL-2007, Popescu-Nanotechnology-2008}) or, even, to be used as probes of the quantum vacuum (see, for instance, Ref. \cite{Dalvit-PRL-2008}).
In the context of the interaction between a particle and a corrugated surface, nontrivial geometric effects can be predicted when calculations are made beyond the proximity force approximation (PFA)
\cite{Dalvit-PRL-2008, Dalvit-JPA-2008, Dobrich-PRD-2008, Messina-PRA-2009, Moreno-NJP-2010, Reyes-PRA-2010, Moreno-PRL-2010, Bimonte-PRD-2014, Marachevsky-TMP-2015, Bennett-PRA-2015, Buhmann-IJMPA-2016, Nogueira-PRA-2021,Queiroz-PRA-2021}. 
For example, in Ref. \cite{Dalvit-PRL-2008}, it is shown that for an isotropic particle interacting with a grooved surface, the PFA predicts that no lateral force acts on the particle when it is over a plateau of the surface, whereas analyses beyond this approximation reveal the existence of a lateral force.
In the present paper, we consider a single slight protuberance in a perfectly conducting plane, and, by making calculations valid beyond the PFA, we predict that sign inversions and other nontrivial geometric effects related to the lateral vdW force arise when we consider a neutral anisotropic particle, and manipulate the ratio between the characteristic widths of the protuberance and the distance of this particle from the plane.
Moreover, we investigate situations where the sign inversion in the lateral vdW force does not occur for a single protuberance, but arises when two or more protuberances are put together, distinguishing between sign inversions originated by individual or collective effects.

The paper is organized as follows. 
In Sec. \ref{sec-revisited}, we discuss the approach to compute the vdW interaction between a neutral particle and a general grounded conducting corrugated surface. 
In Sec. \ref{sec:gauss}, we apply our formulas to the case of a plane with a single Gaussian protuberance.
In Sec. \ref{sec-singlestrip}, we apply our formulas to the case of a plane with a single rectangular strip and generalize this situation to the case of two or more rectangular strips.
In Sec. \ref{sec-final}, we present our final remarks.

%%%%%%%%%%%%%%%%%%%%%%%%%%%%%%%%%%%%%%%%%%%%%%%%%%%%%%%%%%%%%%%%%%%%%%%%%%%%%%%%%%%%%%%%%
%
\section{Model and approach}
\label{sec-revisited}
 
\begin{figure}[h]
\centering
\epsfig{file=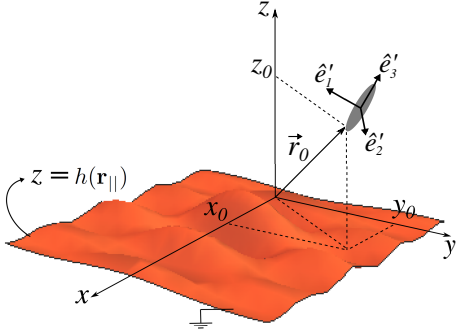,  width=0.7 \linewidth}
\caption{
Illustration of a neutral polarizable anisotropic particle, arbitrarily oriented in space, located at ${\bf r}_{0}={\bf r}_{0\parallel}+z_0\hat{{\bf z}}$ (with $z_0>0$), interacting with a general grounded conducting corrugated surface, whose corrugation profile is described by $z=h(\textbf{r}_\parallel)$.
The unit vectors $\hat{\mathbf{e}}_1^{\prime}$, $\hat{\mathbf{e}}_2^{\prime}$ and $\hat{\mathbf{e}}_3^{\prime}$  
are those pointing to the directions of the principal axes of the particle, in whose directions the particle presents the polarizabilities $\alpha_1(\omega)$, $\alpha_2(\omega)$, and $\alpha_3(\omega)$, respectively.
}
\label{fig:particula-superficie-geral}
\end{figure}

Let us start considering an anisotropic polarizable particle characterized by a frequency dependent polarizability tensor $\boldsymbol{\alpha}$ given by (see, for instance, Ref. \cite{Thiyam-PRA-2015})
$\boldsymbol{\alpha}(\omega)=\alpha_1(\omega)\hat{\mathbf{e}}_1^{\prime}\hat{\mathbf{e}}_1^{\prime}
+\alpha_2(\omega)\hat{\mathbf{e}}_2^{\prime}\hat{\mathbf{e}}_2^{\prime}
+\alpha_3(\omega)\hat{\mathbf{e}}_3^{\prime}\hat{\mathbf{e}}_3^{\prime}$,
where $\hat{\mathbf{e}}_1^{\prime}$, $\hat{\mathbf{e}}_2^{\prime}$ and $\hat{\mathbf{e}}_3^{\prime}$ are unit vectors pointing in the directions of the principal axes of the particle (along any one of them, an applied electric field induces a dipole moment in the same direction \cite{Feynman-Lectures-vol-2}).
Denoting the Euler angles by $(\phi,\theta,\psi)$, 
according to the convention usually adopted in quantum mechanics, we have
$\hat{\mathbf{e}}_i=\sum_j R_{ij}\hat{\mathbf{e}}^{\prime}_j$,
where $R_{ij}$ are the elements of the Euler rotation matrix $R(\phi,\theta,\psi)$ \cite{Sakurai-QM-1994, Ballentine-QM-2014},
$\hat{\mathbf{e}}_1=\hat{{\bf x}}$, $\hat{\mathbf{e}}_2=\hat{{\bf y}}$, and 
$\hat{\mathbf{e}}_3=\hat{{\bf z}}$ are unit vectors related to the laboratory coordinate system $xyz$.
The particle is located at ${\bf r}_{0}={\bf r}_{0\parallel}+z_0\hat{{\bf z}}$ (with $z_0>0$
and ${\bf r}_{0\parallel}=x_0\hat{{\bf x}}+y_0\hat{{\bf y}}$),
above a grounded conducting corrugated surface described by $z=h(\mathbf{r}_{\parallel})$, 
with $h({\bf r}_{||})$ describing a general small modification $[\text{max}|h({\bf r}_{||})|=a\ll z^\prime]$ to a grounded planar conducting surface at $z=0$ (see Fig. \ref{fig:particula-superficie-geral}).

To investigate the vdW interaction energy $U_\text{vdW}$ between the particle and the corrugated surface, 
we take as basis the analytical perturbative approach presented by us in Ref. \cite{Nogueira-PRA-2021},
according to which $U_\text{vdW}\approx U^{(0)}_{\text{vdW}} + U^{(1)}_{\text{vdW}}$, where 
$U^{(0)}_{\text{vdW}}$ is the vdW potential for the case of a grounded conducting plane
\cite{Lennard-Jones-TransFarSoc-1932}, 
and $U^{(1)}_{\text{vdW}}$, the first-order correction to $U^{(0)}_{\text{vdW}}$ due to the surface corrugation, is given by
\begin{equation}
U^{(1)}_{\text{vdW}}(\mathbf r_0)= -\frac{\hbar a}{64\pi^2\epsilon_0z_0^4}\sum_{i,j}{\cal K}_{ij}({\bf r}_0,h) { \int_{0}^{\infty}d\xi\,\alpha_{ij}\left(i\xi\right)}
\end{equation}
where
$\alpha_{ij}$ are the components of the polarizability
tensor in the laboratory system, the functions ${\mathcal{K}}_{ij}$ are written as
\begin{equation}
{\mathcal{K}}_{ij}({\bf r}_0,h)=
\frac{1}{a}\int\frac{d^{2}{\bf q}}{(2\pi)^{2}}\tilde{h}({\bf q})e^{i{\bf q}\cdot{\bf r}_{0\parallel}}\mathcal{J}_{ij}(z_{0}{\bf q}),
\label{kappa-geral}
\end{equation}
with the functions $\mathcal{J}_{ij}(\mathbf{u})=\mathcal{J}_{ji}(\mathbf{u})$ given by
\begin{align}
	\nonumber
	\mathcal{J}_{kl} & =\delta_{kl}\frac{3}{8}| {{\bf u}}|^{3}K_{3}(| {{\bf u}}|)-\frac{3}{8} u_{k}u_{l}| {{\bf u}}|^{2}K_{2}(| {{\bf u}}|) \;\; (k,l=x,y), \\
	\mathcal{J}_{zz}& =\left(2+\frac{3}{8}| {{\bf u}}|^{2}\right) |{{\bf u}}|^{2}K_{2}(| {{\bf u}}|)+\frac{1}{4}| {{\bf u}}|^{3}K_{3}(| {{\bf u}}|),\\
	\nonumber
	\mathcal{J}_{kz} & =i{u}_{k}| {{\bf u}}|^{2}K_{2}\left(| {{\bf u}}|\right)-\frac{3i}{8} {u}_{k}| {{\bf u}}|^{3}K_{3}\left(| {{\bf u}}|\right) \;\; (k=x,y), 
\end{align}
where $K_{2}$ and $K_{3}$ are modified Bessel functions of the second kind.
Note that, 
$a/z_0 \ll 1$ is our perturbative
parameter and ${\cal K}_{ij}$ are dimensionless functions storing information
on the corrugation.

Since we are interested in
a particle arbitrarily oriented in space, using the Euler angles we write
$
\int_{0}^{\infty}d\xi\,\alpha_{ij}\left(i\xi\right)=
R(\phi,\theta,\psi)A R^{-1}(\phi,\theta,\psi),
$
where $A=\int_{0}^{\infty} d\xi\,\text{diag}[\alpha_{1}\left(i\xi\right),\alpha_{2}\left(i\xi\right),\alpha_{3}\left(i\xi\right)]$.
Motivated by the discussion in Ref. \cite{Lewis-JPhysChemA-2000}, we write
$
A=\gamma_{\text{iso}}\Pi(\gamma_s,\gamma_a)
$
where $\Pi(\gamma_s,\gamma_a)=I+\gamma_s M_s+\gamma_a M_a$, 
$I$ is the $3\times 3$ identity matrix, $M_{s}=\text{diag}(-1,-1,2)$,
$M_{a}=\text{diag}(-3,3,0)$,
and we introduced the parameters $\gamma_{\mathrm{iso}}$, $\gamma_s$, and $\gamma_a$,
which characterize, in a convenient manner, the particle anisotropy, and are given by
\begin{equation}
\gamma_{\mathrm{iso}}=\frac{\text{Tr}(A)}{3},\:\gamma_{s}=\frac{[A_{33}-\frac{(A_{22}+A_{11})}{2}]}{3\gamma_{\mathrm{iso}}},\:\gamma_{a}=\frac{\frac{(A_{22}-A_{11})}{2}}{3\gamma_{\mathrm{iso}}},
\end{equation}
assuming that the principal axes of the particle have been enumerated in such a way that $A_{11}\leq A_{22}\leq A_{33}$.
The $\gamma$ parameters are such that $0\leq \gamma_s<1$, and $0\leq \gamma_a\leq \min(\gamma_s, 1-\gamma_s)$.
The parameter $\gamma_s$ measures the degree of particle anisotropy, in the sense that $\gamma_s= 0$ means an isotropic particle, whereas the greater the value of $\gamma_s$, the greater the difference between the polarizability along $\hat{\mathbf{e}}_3^{\prime}$ and the others along $\hat{\mathbf{e}}_1^{\prime}$ and $\hat{\mathbf{e}}_2^{\prime}$.
On the other hand, $\gamma_a$ is an adittional measure of the particle anisotropy, by means of the difference between the polarizabilities along $\hat{\mathbf{e}}_1^{\prime}$ and $\hat{\mathbf{e}}_2^{\prime}$.
Taking all this into account, we write 
\begin{equation}
\frac{U_{\mathrm{vdW}}^{(1)}(\mathbf{r}_{0})}{\mathcal{U}(z_{0})}=-\text{Tr}[{\cal K}(\mathbf{r}_{0},h)R(\phi,\theta,\psi)\Pi(\gamma_{s},\gamma_{a})R^{-1}(\phi,\theta,\psi)],
\label{eq:main-rewritten}
\end{equation}
where ${\mathcal{U}(z_0)} = {\hbar \gamma_{\mathrm{iso}}a}/{(64\pi^{2}\epsilon_{0}z_{0}^{4})}$.
The dimensionless ratio 
$U_{\mathrm{vdW}}^{(1)}/{\mathcal{U}(z_0)}$
in Eq. \eqref{eq:main-rewritten} is useful to investigate the behavior of the lateral vdW force for
an isotropic particle and a general anisotropic one, arbitrarily oriented in space, interacting with
a perfectly conducting corrugated surface. 
With this in mind, hereafter we consider the particle kept constrained to move on the plane $z = z_0$.
In addition, a classical counterpart of Eq. \eqref{eq:main-rewritten}, involving a neutral particle with a permanent electric dipole moment, can be obtained in a similar way (see Appendix \ref{app:classical}).

%%%%%%%%%%%%%%%%%%%%%%%%%%%%%%%%%%%%%%%%%%%%%%%%%%%%%%%%%%%%%%%%%%%%%%%
\section{A Gaussian protuberance}
\label{sec:gauss}

We first apply our Eq. \eqref{eq:main-rewritten} to investigate the vdW interaction for the situation of a plane surface with a single slight Gaussian protuberance of height $a$ and width $d$, given by $h(\mathbf{r}_{\parallel})= a\exp[{-\left(|\mathbf{r}_{\parallel}|/d\right)^2}]$. 
For this case, one obtains that 
\begin{equation}
\mathcal{K}_{ij}\left(\frac{{\bf r}_{0\parallel}}{z_{0}},\frac{d}{z_{0}}\right)=\pi\frac{d^{2}}{z_{0}^{2}}\int\frac{d^{2}{\bf u}}{(2\pi)^{2}}e^{-\frac{1}{4}\frac{d^{2}}{z_{0}^{2}}|{\bf u}|^{2}}e^{i{\bf u}\cdot\frac{{\bf r}_{0\parallel}}{z_{0}}}\mathcal{J}_{ij}({\bf u}),
\label{eq:kappa-gaussiana}
\end{equation}
and $U_{\mathrm{vdW}}^{(1)}/\mathcal{U}$ depends on the ratios $x_0/z_0$, $y_0/z_0$ and $d/z_0$.
The geometric effects of the Gaussian protuberance are regulated by the ratio $d/z_0$, so that we are going to investigate how the variation of this ratio affects the behavior of $U_{\mathrm{vdW}}^{(1)}/\mathcal{U}$ in relation to $x_0/z_0$ and $y_0/z_0$.

In the limit of a very smooth Gaussian $(d/z_0 \to \infty)$, one recovers the result characteristic of the PFA \cite{Dalvit-JPA-2008}, and Eq. \eqref{eq:main-rewritten} becomes
\begin{align}
\frac{U_{\mathrm{vdW}}^{(1)}}{\mathcal{U}} & \approx-3e^{-\left(|\mathbf{r}_{\parallel}|/d\right)^{2}} \nonumber \\
& \times\text{Tr}[\text{diag}(1,1,2)R(\phi,\theta,\psi)\Pi\left(\gamma_{s},\gamma_{a}\right)R^{-1}(\phi,\theta,\psi)], \label{eq:U-PFA}
\end{align}
In this context, one can observe that the minimum values of  $U_{\mathrm{vdW}}^{(1)}$ always coincide with the planar coordinates of the Gaussian peak $(|\mathbf{r}_{\parallel}|=0)$, no matter if the particle is isotropic or anisotropic (or how this latter is oriented).
In other words, when the PFA is considered, the prediction is that a particle will always be attracted to the protuberance,
under the action of the lateral vdW force.  	

\begin{figure}[h]
\centering  
\subfigure[]{\label{fig:gaussiana-PFA}\epsfig{file=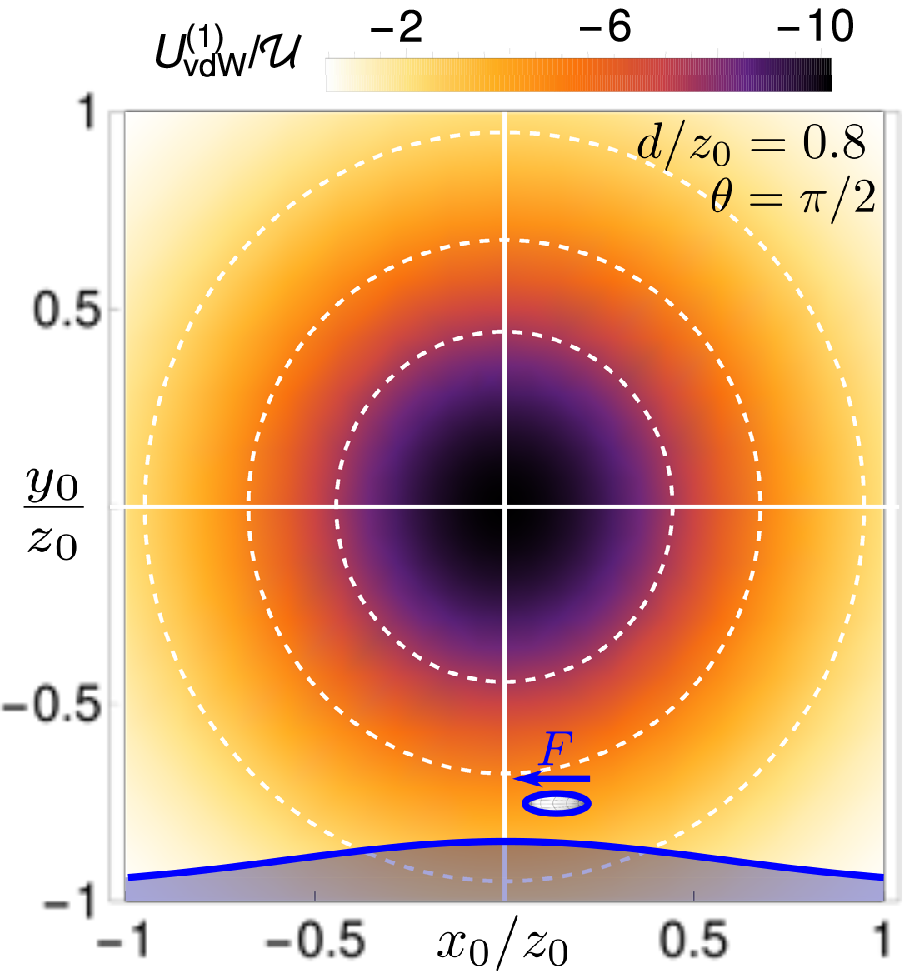, width=0.49 \linewidth}}
\hspace{0mm}
\subfigure[]{\label{fig:gaussiana-x-08}\epsfig{file=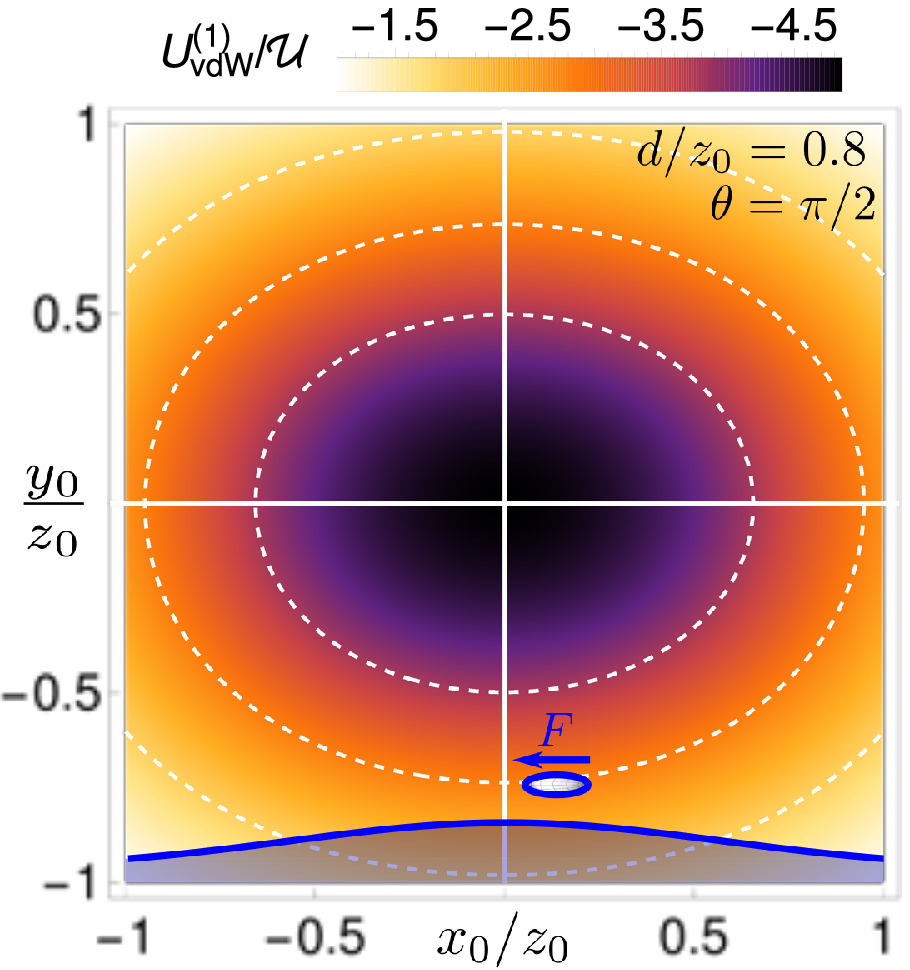, width=0.49 \linewidth}}
\hspace{0mm}
\subfigure[]{\label{fig:gaussiana-x-02}\epsfig{file=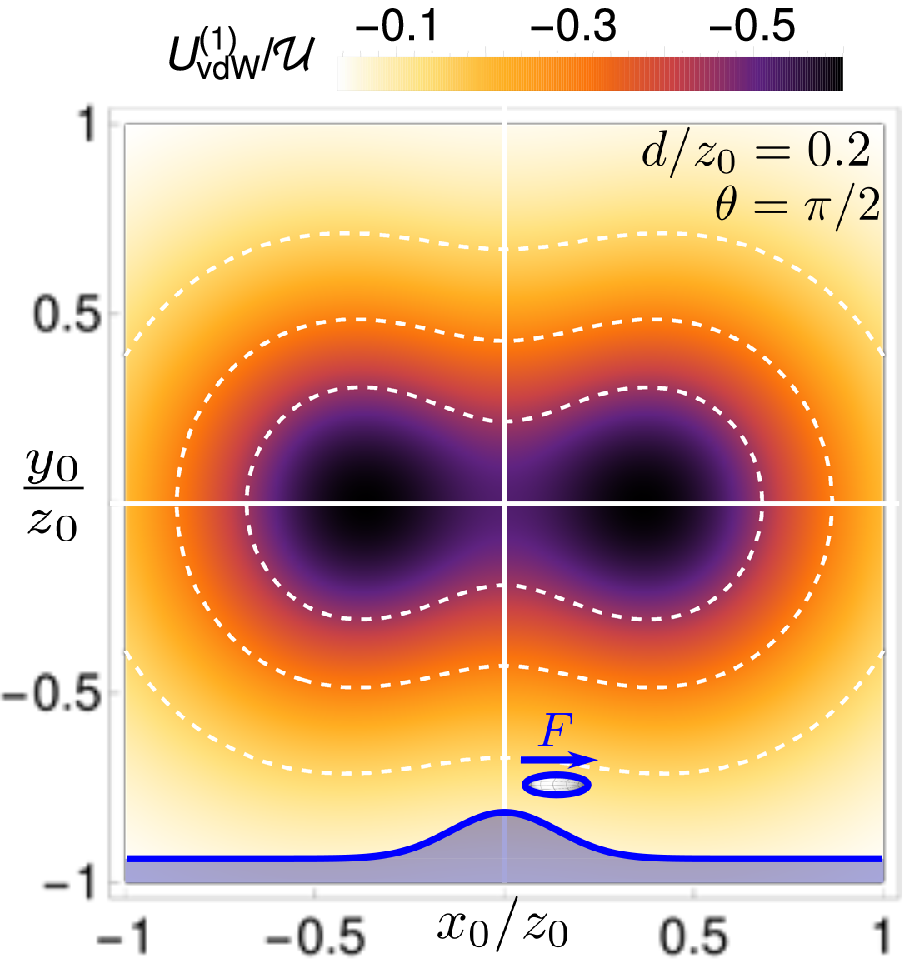, width=0.49 \linewidth}}
\hspace{0mm}
\subfigure[]{\label{fig:gaussiana-orientacao-0}\epsfig{file=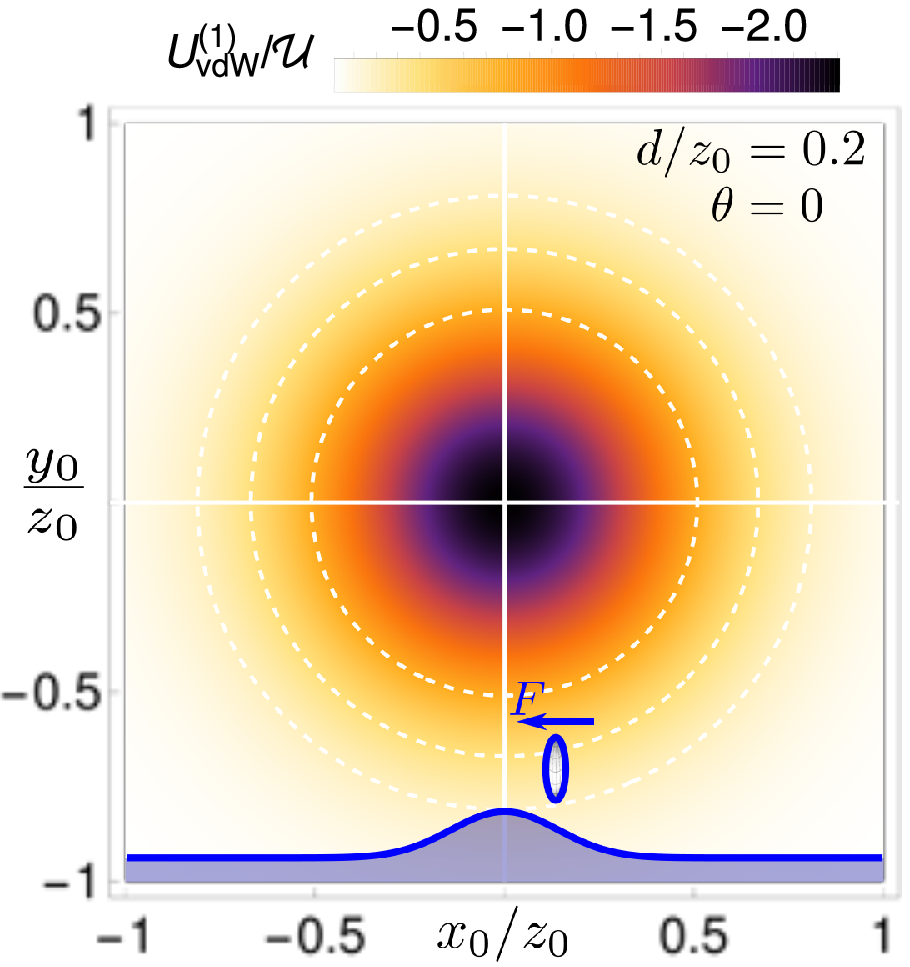, width=0.49 \linewidth}}
\caption{
Behavior of $U_{\mathrm{vdW}}^{(1)}/{\cal{U}}$ versus ${x_0}/{z_0}$ and ${y_0}/{z_0}$, for a particle fixed at $z = z_0$.
In each figure, we consider this particle characterized by $\gamma_{a}=0$ and $\gamma_{s}=0.6$, oriented such that $\phi=0$ and $\psi=0$.
The insets in the bottom of the graphics illustrate the Gaussian profile projected on the 
plane $y=0$, a particle located at $(x_0,0,z_0)$, its orientation and the direction of the lateral vdW force acting on it.
In (a), the PFA is considered. 
In (b) - (d), we consider our Eq. \eqref{eq:main-rewritten}, providing predictions beyond the PFA.
From (b) to (c), the Gaussian width decreases (decreasing the ratio $d/z_0$).
From (c) to (d), there is a change on the particle orientation.
}
\label{fig:regiao-epsilon}
\end{figure}

For a generic value of $d/z_0$,
when considering an isotropic particle  $(\gamma_s = 0)$,
Eq. \eqref{eq:main-rewritten} becomes simply $U_{\mathrm{vdW}}^{(1)}/\mathcal{U}=-\text{Tr}({\cal K})$, and we obtain that $U_{\mathrm{vdW}}^{(1)}$ has a single minimum point at $|\mathbf{r}_{\parallel}|=0$ for any value of $d/z_0$, 
which means that, under the action of the lateral vdW force, an isotropic particle is always attracted to the peak of the Gaussian protuberance.
On the other hand, when we consider an anisotropic particle, our Eq. \eqref{eq:main-rewritten} shows that nontrivial geometric effects arise when the ratio $d/z_0$ decreases.
Hereafter, for simplicity, we focus our analysis on the case of generic particles characterized by $\gamma_{a}=0$ (cylindrically symmetric polarizable particles) and oriented with $\phi=0, \psi=0$ (which means that $\hat{\mathbf{e}}_3^{\prime}$ is on the $xz$-plane).
In this context, let us consider ${d}/{z_0}=0.8$, $\gamma_s = 0.6$, and $\theta=\pi/2$ ($\hat{\mathbf{e}}_3^{\prime}=\hat{\mathbf{x}}$).
When the PFA is used, we obtain that the equipotential lines are circular, and the minimum value of $U^{(1)}_{\text{vdW}}$ is located at the origin, so that the lateral vdW force attracts the particle to the protuberance peak [see Fig. \ref{fig:gaussiana-PFA}].
On the other hand, a better description of the behavior of $U^{(1)}_{\text{vdW}}/{\cal{U}}$ is obtained by using Eq. \eqref{eq:main-rewritten} and shown in Fig. \ref{fig:gaussiana-x-08}, from where one can see that the minimum point of $U^{(1)}_{\text{vdW}}$ remains at the origin, but the equipotential lines are, in fact, ellipsoidal.
Now, let us decrease the ratio $d/z_0$, considering, for example, $d/z_0=0.2$.
In Fig. \ref{fig:gaussiana-x-02}, we show the behavior of  $U^{(1)}_{\text{vdW}}/{\cal{U}}$ for this case, according to Eq. \eqref{eq:main-rewritten}.
One can see that, now,  the energy has two minimum points, and none of them coincide with the peak of the Gaussian.
As a consequence of this change in the minimum points, one has a considerable change in the shape of the equipotential lines, and the particle can feel a lateral vdW force which moves it away from the peak of the protuberance.
Thus, when a particle is slightly dislocated from the 
point $(0,0,z_0)$,
along the $x$-axis, the lateral force points to opposite directions when comparing the situations with ${d}/{z_0}=0.8$ [Fig. \ref{fig:gaussiana-x-08}] and ${d}/{z_0}=0.2$ [Fig. \ref{fig:gaussiana-x-02}].
This sign inversion in the lateral vdW force is therefore
a nontrivial geometric effect regulated by the ratio $d/z_0$.
Moreover, this effect is affected by the particle orientation.
From the case shown in Fig. \ref{fig:gaussiana-x-02}, for example, if we change the particle orientation to $\theta=0$ $(\hat{\mathbf{e}}_3^{\prime} = \hat{\mathbf{z}})$, one sees that the energy has only one minimum point located at the origin and the equipotencial lines are circular [see Fig. \ref{fig:gaussiana-orientacao-0}], so that the lateral vdW force leads the particle to the Gaussian peak, independently of the ratio $d/z_0$ (a discussion of cases where $0 < \theta < \pi/2$ can be found in Appendix \ref{app:orientacao}).

\begin{figure}[h]
\centering
\subfigure[]{\label{fig:gamma-delta-gaussiana}\epsfig{file=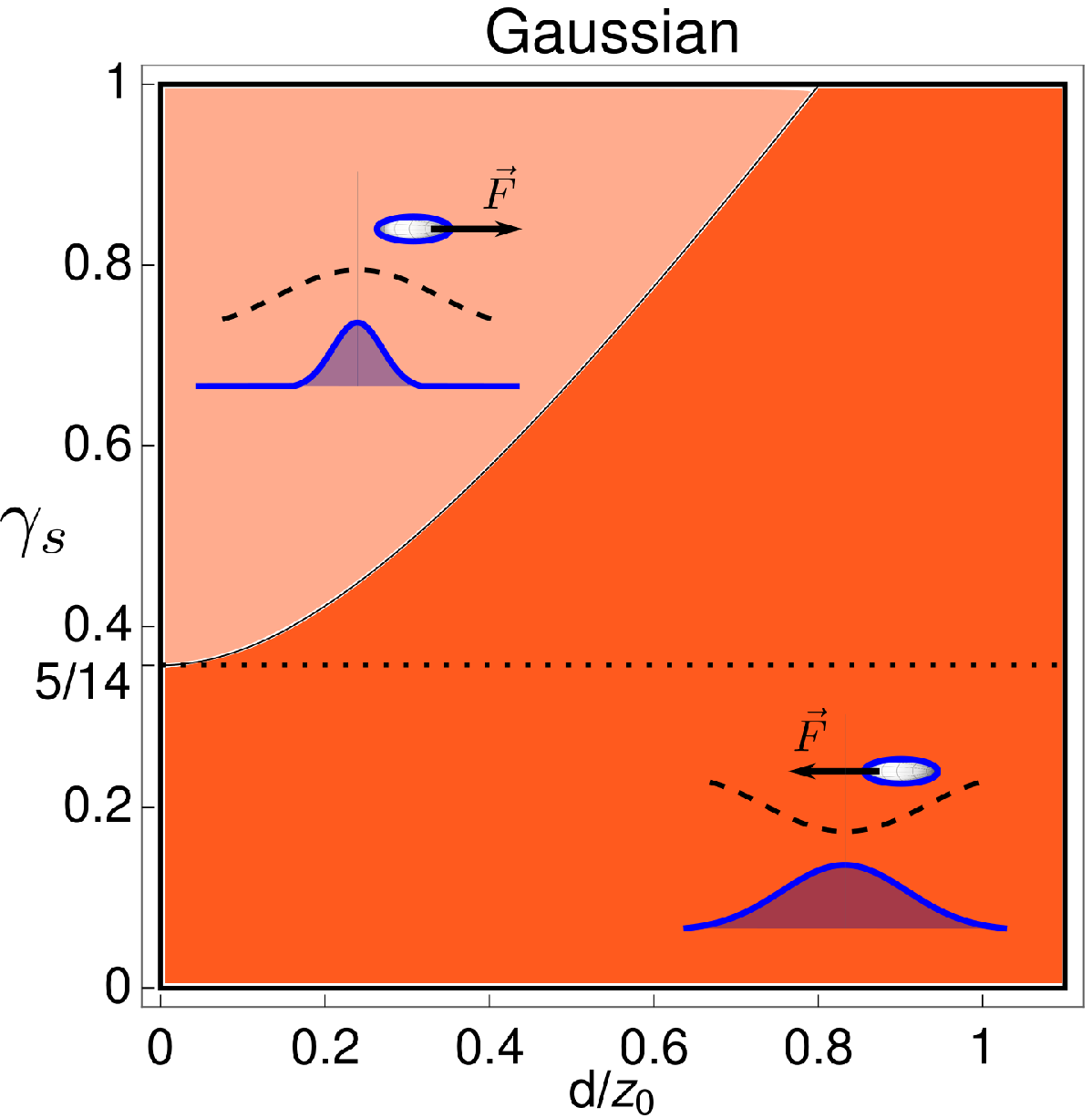, width=0.48 \linewidth}}
\hspace{1mm}
\subfigure[]{\label{fig:gamma-delta-strip}\epsfig{file=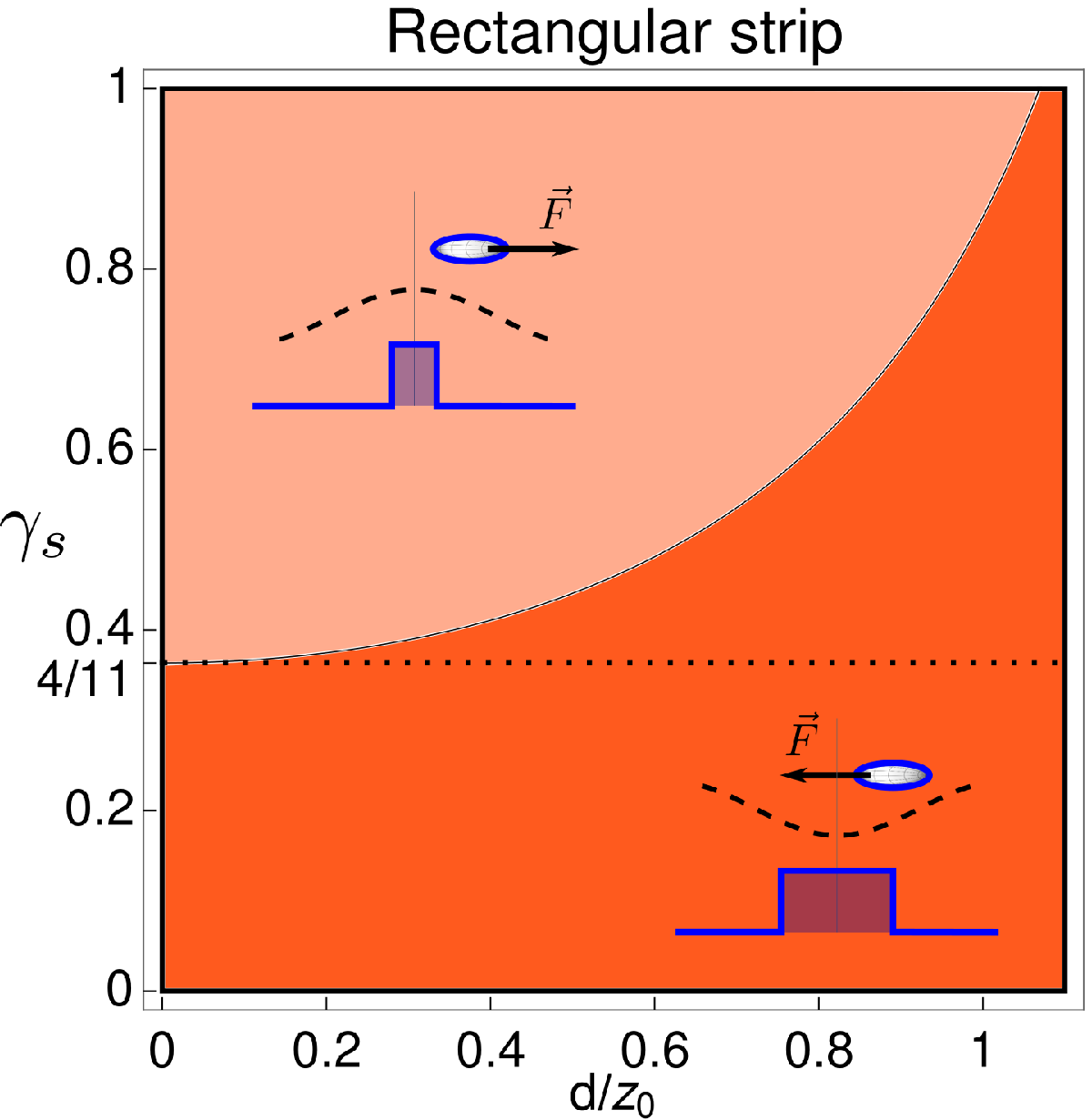, width=0.48 \linewidth}}
\caption{
For a particle characterized by $\gamma_{a}=0$ and oriented with $\phi=0,\theta=\pi/2,\psi=0$, it is shown the configurations of $\gamma_{s}$ and $d/z_0$ for which the center of the protuberance is a minimum (dark region) or a maximum (lighter region) point of $U_{\mathrm{vdW}}^{(1)}/{\cal{U}}$, as illustrated in the insets.
In (a), we show this for a Gaussian protuberance, whereas in (b), for a rectangular strip.
The dotted lines highlight the values of $\gamma_{s}$ [$5/14$ in (a) and $ 4/11$ in (b)], below which there is no value of $d/z_0$ that makes possible a transition from the dark to the lighter region, so that there is no possibility of a sign inversion in the lateral force.
}
\label{fig:gamma-delta}
\end{figure}

All these nontrivial effects are also influenced by the particle anisotropy.
We consider a generic particle characterized by $\gamma_a=0$ and an arbitrary value for $\gamma_s$, oriented with $\phi=\psi=0$ and $\theta=\pi/2$. 
In Fig. \ref{fig:gamma-delta-gaussiana} we show the configurations of $\gamma_{s}$ and $d/z_0$ for which the center of the protuberance is a minimum (dark region) or a maximum (lighter region) point of $U_{\mathrm{vdW}}^{(1)}/{\cal{U}}$.
The sign inversion in the lateral force is possible when there is a transition from a configuration in the dark region to another in the lighter one.
For instance, when we consider $\gamma_{s} = 0.6$, and change the ratio ${d}/{z_0}$ from ${d}/{z_0}=0.8$ to ${d}/{z_0}=0.2$, we have a transition from the dark to the lighter region in Fig. \ref{fig:gamma-delta-gaussiana}, which corresponds to the transition from Figs. \ref{fig:gaussiana-x-08} to \ref{fig:gaussiana-x-02}.
On the other hand, when considering $\gamma_s < 5/14$, there is no change of the ratio ${d}/{z_0}$ that makes possible a transition from the dark to the lighter region, so that there is no possibility of a sign inversion in the lateral force.
In addition, the change from the dark to the lighter region in Fig. \ref{fig:gamma-delta-gaussiana} can also be obtained by considering a fixed value for $d$, and increasing $z_0$, so that as $z_0$ increases, the more evident are the effects predicted here.
In this way, even if $z_0$ becomes large enough for the retardation effects become relevant (Casimir-Polder regime of the interaction), we expect that a sign inversion in the lateral force will still occur.

\begin{figure}[h]
\centering  
\subfigure[]{\label{fig:buraco-PFA}\epsfig{file=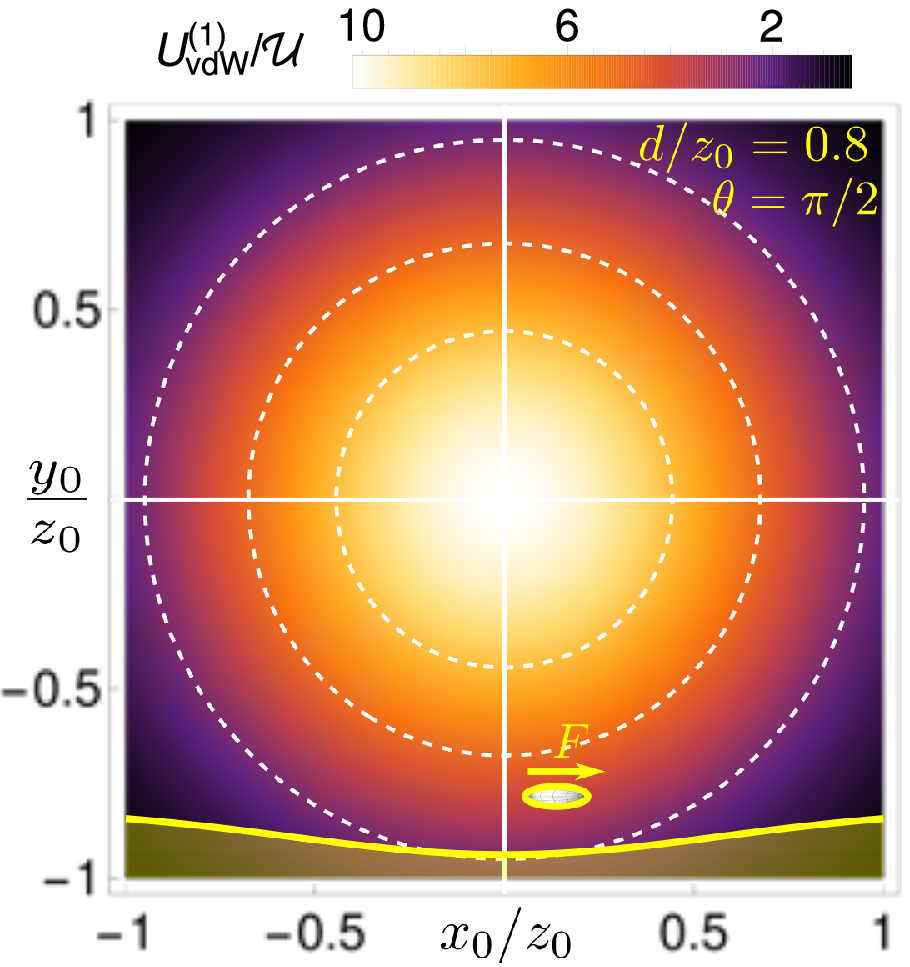, width=0.49 \linewidth}}
\hspace{0mm}
\subfigure[]{\label{fig:buraco-x-08}\epsfig{file=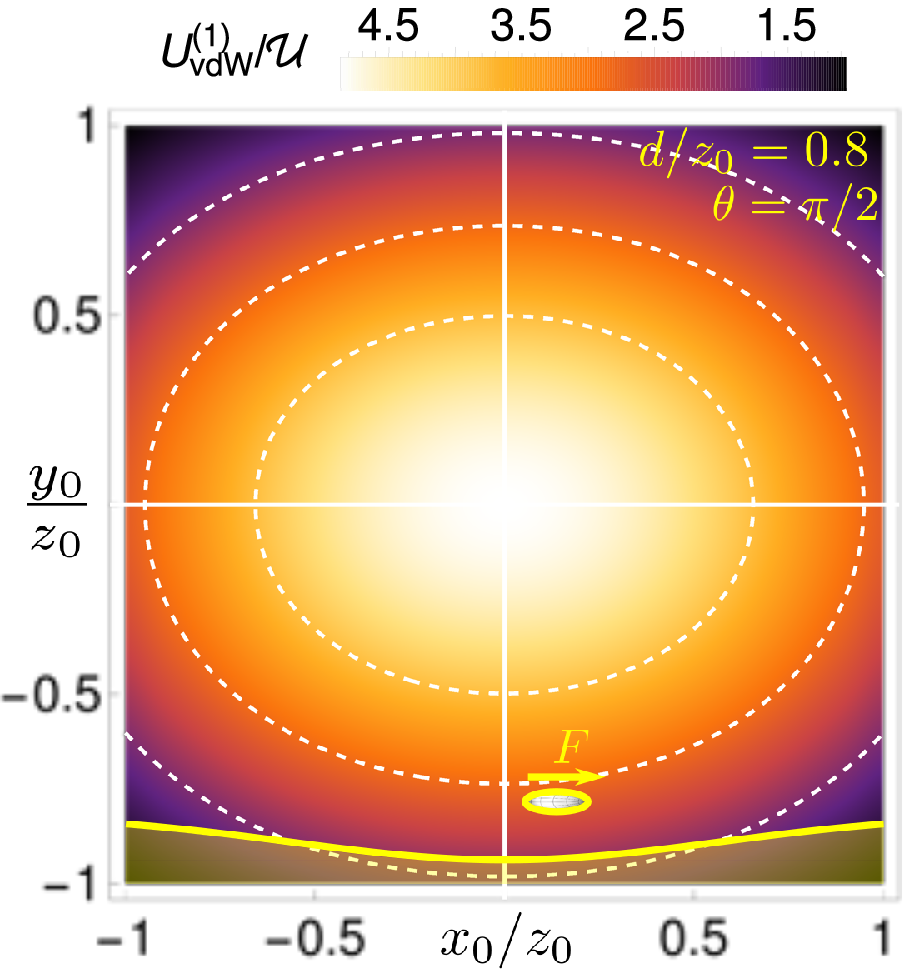, width=0.49 \linewidth}}
\hspace{0mm}
\subfigure[]{\label{fig:buraco-x-02}\epsfig{file=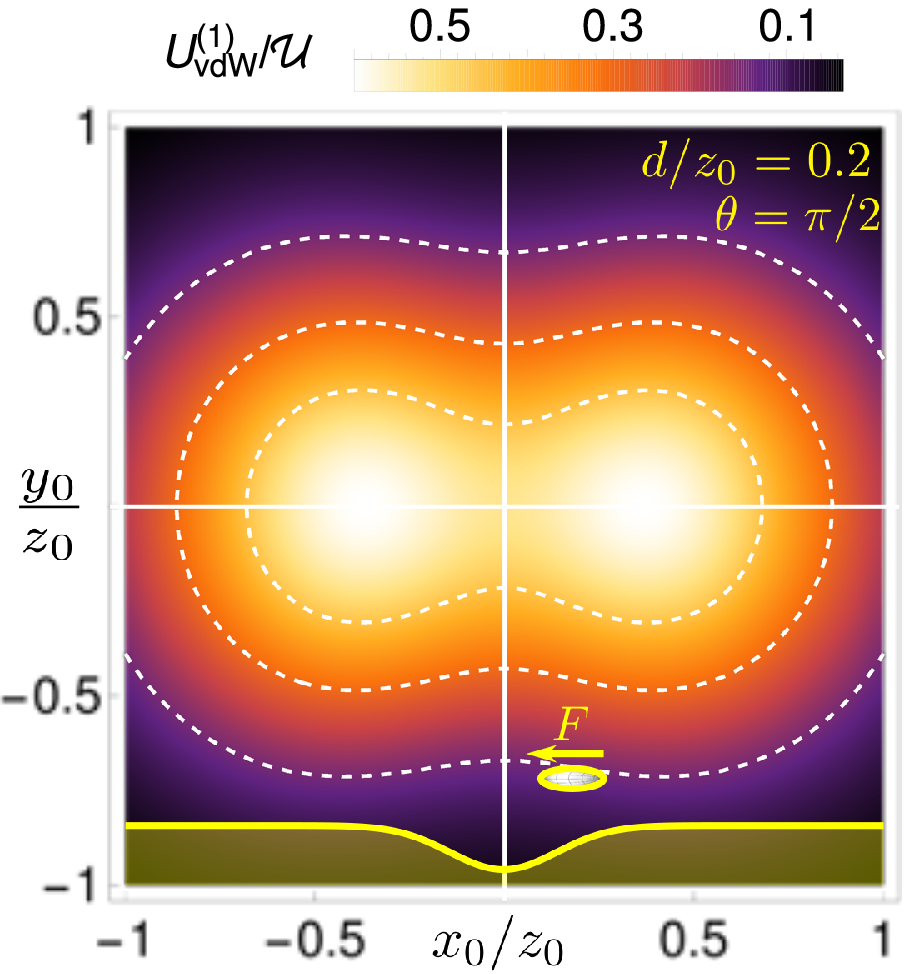, width=0.49 \linewidth}}
\hspace{0mm}
\caption{
Behavior of $U_{\mathrm{vdW}}^{(1)}/{\cal{U}}$ versus ${x_0}/{z_0}$ and ${y_0}/{z_0}$, for a particle fixed at $z = z_0$ interacting with a plane with a Gaussian hole of width $d$.
In each figure, we consider this particle characterized by $\gamma_{a}=0$ and $\gamma_{s}=0.6$, oriented such that $\phi=0$ and $\psi=0$.
The insets in the bottom of the graphics illustrate the Gaussian hole projected on the plane $y=0$, a particle located at $(x_0,0,z_0)$, its orientation and the direction of the lateral vdW force acting on it.
In (a), the PFA is considered, whereas in (b) and (c), we consider predictions beyond this approximation.
From (b) to (c), the width of the Gaussian hole decreases (decreasing the ratio $d/z_0$).
Note that, when a particle is slightly dislocated from the point $(0,0,z_0)$, along the $x$-axis, the lateral force points to opposite directions when comparing these situations.
}
\label{fig:buraco}
\end{figure}

When, instead of considering a protuberance $z=h(x)\geq0$ in Eq. \eqref{eq:main-rewritten}, we consider a hole $[h(x)\leq 0]$, the sign of $U_{\mathrm{vdW}}^{(1)}$ is inverted,
which means that the minimum points of $U_{\mathrm{vdW}}^{(1)}$ for a protuberance become maximum points for a hole, and vice-versa.
In this way, for a Gaussian hole, with characteristic width $d$, when the PFA formula is applied, the particle is always moved away from the hole, no matter if it is isotropic or anisotropic [see Fig. \ref{fig:buraco-PFA}].
On the other hand, using Eq. \eqref{eq:main-rewritten}, while an isotropic particle is always moved away from the hole, independently of the ratio $d/z_0$, an anisotropic particle can be attracted to the hole when the ratio $d/z_0$ is decreased [see Figs. \ref{fig:buraco-x-08} and \ref{fig:buraco-x-02}].
%%%%%%%%%%%%%%%%%%%%%%%%%%%%%%%%
%
%
%

\begin{figure}[h]
\centering  
\subfigure[]{\label{fig:armadilha-08}\epsfig{file=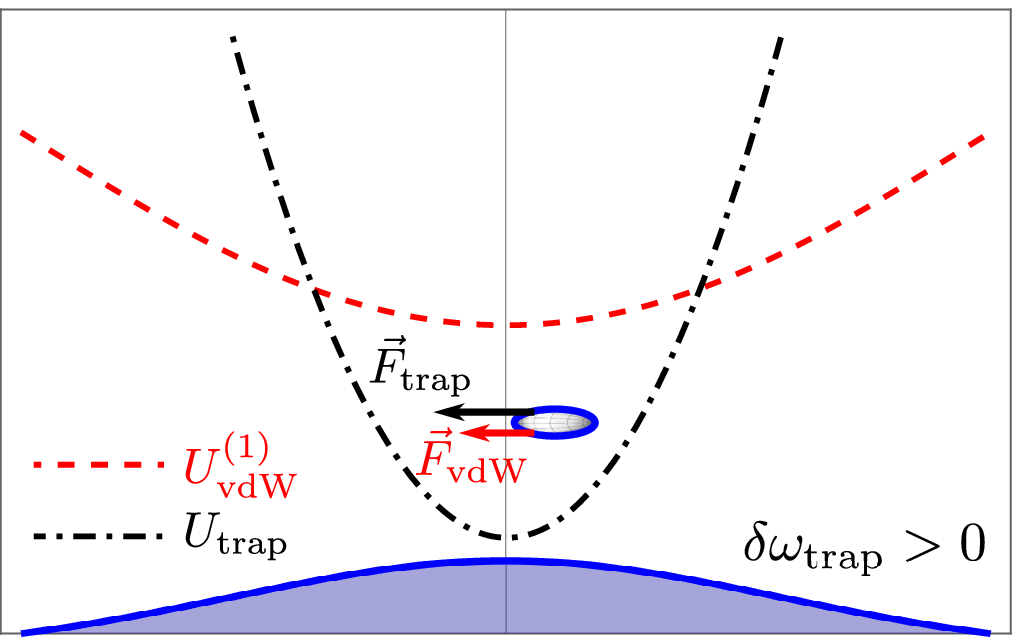, width=0.49 \linewidth}}
\hspace{0mm}
\subfigure[]{\label{fig:armadilha-02}\epsfig{file=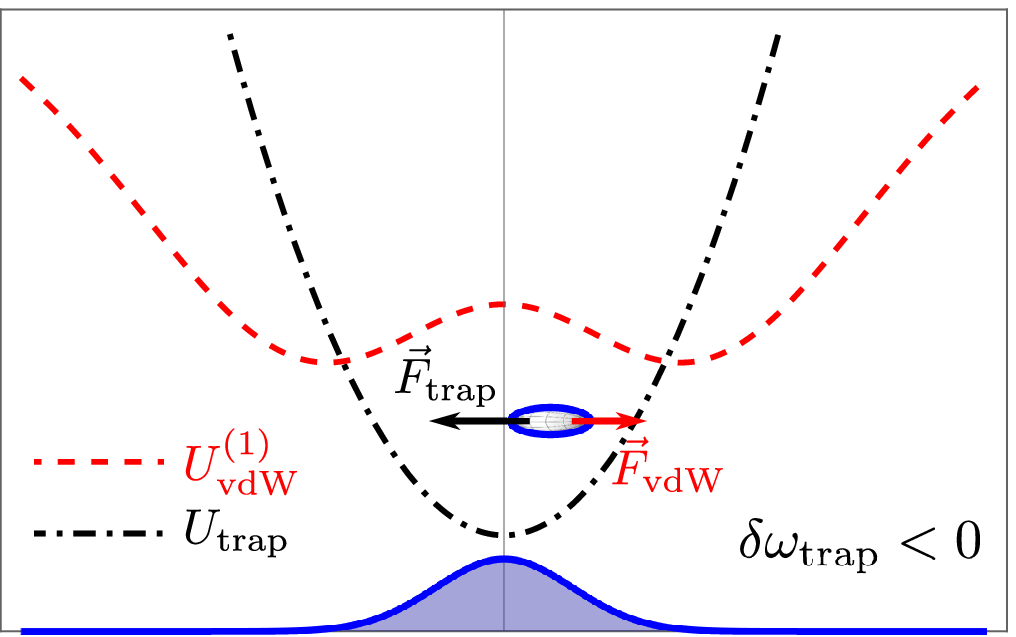, width=0.49 \linewidth}}
\caption{
In (a), where the peak of the protuberance coincides with the minimum point of $U_\text{vdW}^{(1)}$, an experimental apparatus would detect $\delta\omega_\text{trap}>0$.
In (b), the peak of the protuberance coincides with a local maximum point of $U_\text{vdW}^{(1)}$, and for this case one has $\delta\omega_\text{trap}<0$.
Therefore, the sign inversion in the lateral force [illustrated in Figs. \ref{fig:gaussiana-x-08} and \ref{fig:gaussiana-x-02}] manifests, in this scenario, as a sign inversion in the frequency deviation $\delta\omega_\text{trap}$.
}
\label{fig:armadilha}
\end{figure}

The detection of such geometric effects on the lateral vdW force could be done, for example, by trapping a particle with mass $m$, near a plane with a protuberance, and measuring the deviation (due to the protuberance) in the original trap frequency (see, for instance, Refs. \cite{Buhmann-DispersionForces-I, Dalvit-PRL-2008, Nogueira-PRA-2021}).
In the present context, let us consider the situations described in Figs. \ref{fig:gaussiana-x-08} and \ref{fig:gaussiana-x-02} and that the particle is constrained to oscillate only along the $x$-axis, trapped by a harmonic potential $U_\text{trap}(x)$ with equilibrium point at $x_0=0$ (see Fig. \ref{fig:armadilha}).
In the absence of the protuberance, the particle would oscillate around $x_0=0$ with a frequency $\omega_\text{trap}$. 
However, the presence of the protuberance modifies this oscillation frequency to a new value $\omega_\text{trap}^\prime=\sqrt{\omega_\text{trap}^2+m^{-1}\left[\partial^2U_\text{vdW}^{(1)}/\partial x_0^2\right]_{x_0=0}}$, resulting in a frequency deviation 
$\delta\omega_\text{trap}\equiv \omega_\text{trap}^\prime-\omega_\text{trap}$.
When the peak of the protuberance coincides with the minimum point of $U_\text{vdW}^{(1)}$, an experimental apparatus would detect $\delta\omega_\text{trap}>0$ [see Fig. \ref{fig:armadilha-08}].
Otherwise, when the peak of the protuberance coincides with a local maximum point of $U_\text{vdW}^{(1)}$, one has $\delta\omega_\text{trap}<0$ [see Fig. \ref{fig:armadilha-02}].
Therefore, the sign inversion in the lateral force [illustrated in Figs. \ref{fig:gaussiana-x-08} and \ref{fig:gaussiana-x-02}] manifests, in this scenario, as a sign inversion in the frequency deviation $\delta\omega_\text{trap}$.
In addition, as discussed in Refs. \cite{Eberlein-PRA-2007, Nogueira-PRA-2021, Queiroz-PRA-2021}, when the reflectivity of a surface is not perfect, the vdW interaction differs by a numerical factor from the one calculated considering an ideal surface.
Thus, we expect that the sign inversion in the lateral vdW force, and its manifestation as a sign inversion in $\delta\omega_\text{trap}$, manifests itself even when considering nonideal materials.
%

%%%%%%%%%%%%%%%%%%%%%%%%%%%%%%%%%%%%%%%%%%%%%%%%%%%%%%%%%%%%%%%%%%%%%
\section{Rectangular protuberances and gratings}
\label{sec-singlestrip}

\begin{figure}[h]
\centering
\epsfig{file=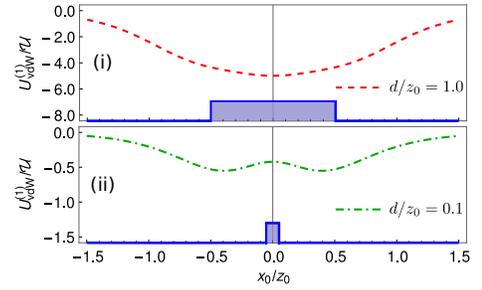, width=.7 \linewidth}
\caption{
Behavior of $U_{\mathrm{vdW}}^{(1)}/{\cal{U}}$ versus ${x_0}/{z_0}$, for a particle fixed at $z = z_0$, and oriented with $\phi=0,\theta=\pi/2,\psi=0$ (in this case, the vector  $\hat{\mathbf{e}}_3^{\prime}$ coincides with $\hat{\mathbf{x}}$).
From top to bottom figure, we change the ratio ${d}/{z_0}$ from ${d}/{z_0}=1$ (dashed line) to ${d}/{z_0}=0.1$ (dot-dashed line).
In these figures, the surface profiles are represented by the solid lines.
Note that, from top to bottom figure, we have a change in the minimum points, so that a particle slightly dislocated from the origin feels a sign inversion in the lateral force.
}\label{fig:largura}
\end{figure}

Since Eq. \eqref{eq:main-rewritten} can be applied to any surface profile, let us apply it to other surface geometries to investigate the generality of the effects discussed so far.
When considering a rectangular strip of height $a$ and width $d$, described by $h(\mathbf{r}_{||})=a \left[\Theta\left(x+\frac{d}{2}\right) - \Theta\left(x-\frac{d}{2}\right)\right]$, where $\Theta$ is the Heaviside step function, we obtain that 
\begin{equation}
	\mathcal{K}_{ij}\left(\frac{x_{0}}{z_{0}},\frac{d}{z_{0}}\right)=\frac{3}{16}[f_{ij}\left(\frac{x_0}{z_0}+\frac{1}{2}\frac{d}{z_0}\right)-f_{ij}\left(\frac{x_0}{z_0}-\frac{1}{2}\frac{d}{z_0}\right)], \label{eq:kappa-strip}
\end{equation}
where
\begin{align}
	f_{xx}(u) & =\frac{u^{3}(8u^{4}+28u^{2}+35)}{(u^{2}+1)^{7/2}}, \nonumber \\
	f_{yy}(u) & =\frac{u(8u^{4}+20u^{2}+15)}{(u^{2}+1)^{5/2}}, \nonumber \\
	f_{zz}(u) & =\frac{u(16u^{6}+56u^{4}+66u^{2}+41)}{(u^{2}+1)^{7/2}},\\
	f_{xz}(u) & =\frac{8u^{2}-7}{(u^{2}+1)^{7/2}}, \nonumber \\
	f_{xy}(u) & =f_{yz}(u)=0, \nonumber 
\end{align}
and $U_{\mathrm{vdW}}^{(1)}/\mathcal{U}$ depends on the ratios $x_0/z_0$ and $d/z_0$.
When the PFA is applied, the prediction is that a particle above the strip feels no lateral vdW force.
On the other hand, according to our Eq. \eqref{eq:main-rewritten}, while an isotropic particle is, in fact, always attracted to the center of the strip (in agreement with Ref. \cite{Dalvit-PRL-2008}), when an anisotropic one is considered,
a sign inversion in the lateral vdW force can occur as $d/z_0$ decreases (see, for instance, Fig. \ref{fig:largura}).
It is remarkable that when considering $\gamma_s < 4/11$ there is no change of the ratio ${d}/{z_0}$ that makes possible a sign inversion in the lateral vdW force [see Fig. \ref{fig:gamma-delta-strip}].

\begin{figure}[h]
\centering
\epsfig{file=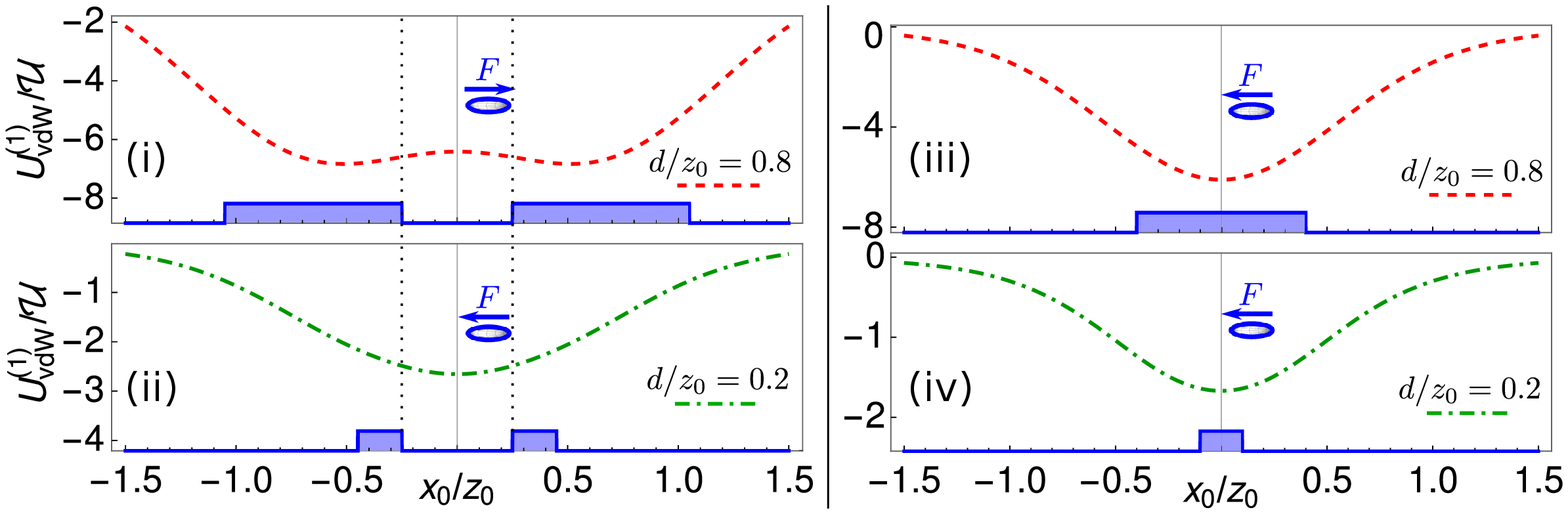, width=1 \linewidth}
\caption{
Behavior of the ratio $U_{\mathrm{vdW}}^{(1)}/{\cal{U}}$ versus ${x_0}/{z_0}$, for a particle fixed at $z = z_0$, with $\gamma_{a}=0$, $\gamma_{s}=0.2$ and oriented with $\phi=0,\theta=\pi/2,\psi=0$.
In (i) and (ii), we consider two strips of width $d$, distant $L$ from each other, with $L/z_0 = 0.5$.
In (iii) and (iv), we consider a single strip of width $d$. 
In these figures, the surface profiles are represented by the solid lines.
From top to bottom figures, we change the ratio ${d}/{z_0}$ from ${d}/{z_0}=0.8$ (dashed line) to ${d}/{z_0}=0.2$ (dot-dashed line).
Note that, from (i) to (ii), we have a change in the minimum points, so that a particle slightly dislocated from the origin feels a sign inversion in the lateral force.
On the other hand, due to the considered value for $\gamma_{s}$, 
from (iii) to (iv), we do not have a change in the minimum point of the energy.
}
\label{fig:largura-strips}
\end{figure}

Sign inversions in the lateral force also occur when we consider more than one protuberance in the plane, but a relevant difference can be observed when we compare this case involving a collective of protuberances with that of a single protuberance. 
To investigate this, let us consider a particle characterized by $\gamma_{s} = 0.2$ and oriented with $\theta=\pi/2$, interacting with two rectangular strips of height $a$, width $d$, and distant by $L$ from each other, which are described by 
$h\left(\mathbf{r}_{||}\right)= a\left[\Theta\left(x+L+\frac{d}{2}\right)-\Theta\left(x+\frac{L}{2}\right)\right] + a\left[\Theta\left(x-\frac{L}{2}\right) - \Theta\left(x-L-\frac{d}{2}\right)\right]$.
According to Eq. \eqref{eq:main-rewritten}, when we have, for instance, $L/z_0 = 0.5$, we obtain that for $d/z_0 = 0.8$ the minimum points of the energy are over the strips [see Fig. \ref{fig:largura-strips}(i)], whereas when $d/z_0 = 0.2$ the energy has only one minimum point located at the middle point between the strips [Fig. \ref{fig:largura-strips}(ii)].
As a consequence of this change in the minimum points, for a particle slightly dislocated from the origin, along the $x$-axis, the lateral force points to opposite directions when comparing both situations
[Figs. \ref{fig:largura-strips}(i) and \ref{fig:largura-strips}(ii)].
Although we have a sign inversion in this case of two strips, if we had considered a single strip, one would see that the change from $d/z_0 = 0.8$ to $d/z_0 = 0.2$ does not change the minimum point of the energy, which remains at the center of the protuberance [see Figs. \ref{fig:largura-strips}(iii) and \ref{fig:largura-strips}(iv)].
This occurs because we are considering $\gamma_s=0.2$, a value for which, in the presence of a single protuberance, there is no possibility of a sign inversion in the lateral force, since it is less than $4/11$. 
%[see Fig. \ref{fig:gamma-delta-strip}].
%
Thus, from all this, one can conclude that the consideration of more than one protuberance is not a mere extension of an individual effect (originated from a single protuberance), but, more than that, there are sign inversions in the lateral force that are induced only by a collective of protuberances.
As the number of rectangular strips increases, so that they tend to form a periodic structure, these collective effects manifest themselves as the peak and valley regimes (see, for instance, Appendix \ref{app:regimes}), which were recently described in the literature \cite{Nogueira-PRA-2021,Queiroz-PRA-2021}.

%%%%%%%%%%%%%%%%%%%%%%%%%%%%%%%%%%%%%%%%%%%%%%%%%%%
\section{Final Remarks}
\label{sec-final}

The interest in knowing how to control the sign of dispersive forces has grown
\cite{Dzyaloshinskii-Lifshitz-Pitaevskii-SPU-1961, Boyer-PhysRev-1968,Feinberg-Sucher-PRA-1970,Boyer-PRA-1974,Kupiszewska-JMP-1993, Farina-Santos-Tort-JPA-2002,Farina-Santos-Tort-AJP-2002,Leonhardt-NJP-2007, Munday-Nature-2009,Levin-PRL-2010,Eberlein-PRA-2011,Milton-JPA-2012,Shajesh-PRA-2012, Hu-JN-2016, Buhmann-IJMPA-2016, Abrantes-PRA-2018,Sinha-PRA-2018,Venkataram-PRA-2020,Maghrebi-PRD-2011, Kenneth-PRL-2002, Adhikari-PRA-2017, Hoye-PRA-2018}.
In this context, we predicted the possibility of sign inversions in the lateral vdW force acting on an anisotropic particle interacting with a perfectly conducting plane containing a single slight protuberance, and showed that these sign inversions can be controlled by suitable manipulations of the ratio between the characteristic widths of the protuberance and the particle-plane distance, opening a new path for investigations of nontrivial aspects of dispersive forces.
Even for situations where the sign inversion does not occur in the presence of a single protuberance, we showed that it can arise when two or more protuberances are put together. 
Thus, we observed a distinction between sign inversions in the lateral vdW force originated by a purely individual effect, from those originated by a collective effect
(such as the peak and valley regimes, recently described in the literature in the context of periodic gratings \cite{Nogueira-PRA-2021,Queiroz-PRA-2021}).

We showed that a sign inversion in the lateral vdW force manifests itself as a sign inversion in the frequency deviation of a particle trapped near a plane with a protuberance.
Thus, the  comprehension of all these nontrivial geometric effects
on the lateral vdW force is relevant to achieve a higher degree of control of the interaction between  a neutral anisotropic polarizable particle and a corrugated surface. 

%%%%%%%%%%%%%%%%%%%%%%%%%%%%%%%%%%%%%%%%%%%%%%%%%%%%%%%%%%%%%%%%%%%%%%%%%%%%%%%%%%%%%%%%%%%
\begin{acknowledgments}
The authors thank Alexandre Costa, Carlos Farina, Stanley Coelho, and Van Sérgio Alves for valuable discussions.
L.Q. and E.C.M.N. were supported by the Coordena\c{c}\~{a}o de Aperfei\c{c}oamento de Pessoal de N\'{i}vel Superior - Brasil (CAPES), Finance Code 001.
\end{acknowledgments}
%

%%%%%%%%%%%%%%%%%%%%%%%%%%%%%%%%%%%%%%%%%%%%%%%%%%%%%%%%%%

\appendix

\section{Classical counterpart of Eq. (5)}
\label{app:classical}

To investigate the existence of sign inversions in the lateral force in a classical context, involving a neutral particle with a permanent electric dipole moment $\bf{p}$,
we again take as basis the perturbative approach shown in Ref. \cite{Nogueira-PRA-2021},
according to which the interaction energy $U_\text{cla}$, between a dipolar particle and a corrugated surface, 
is given by $U_\text{cla}\approx U^{(0)}_{\text{cla}} + U^{(1)}_{\text{cla}}$, where 
$U^{(0)}_{\text{cla}}$ is the interaction energy for the case of a grounded conducting plane
\cite{Jackson-Electrodynamics-1998}, 
and $U^{(1)}_{\text{cla}}$ is given by
$U^{(1)}_{\text{cla}}(\mathbf r_0)= -\frac{a}{z_0}\sum_{i,j}{\cal K}_{ij}({\bf r}_0,h) 
p_{i}p_{j}/{(64\pi\epsilon_{0}z_0^3)}$. 
After manipulations of this formula, we obtain
\begin{equation}
\label{eq:main-rewritten-classical}
\frac{U_{\mathrm{cla}}^{(1)}(\mathbf{r}_{0})}{\mathcal{U}_{\text{cla}}(z_{0})}=-\text{Tr}[{\cal K}(\mathbf{r}_{0},h)R(\phi,\theta,\psi)\Pi(1,0)R^{-1}(\phi,\theta,\psi)],
\end{equation}
where ${\mathcal{U}_{\text{cla}}(z_0)} = a|{\bf{p}}|^2/({192\pi\epsilon_{0}z_{0}^{4}})$.
Note that the behavior of the classical ratio $U_{\mathrm{cla}}^{(1)}/\mathcal{U}_{\text{cla}}$
is the same of the quantum ratio $U_{\mathrm{vdW}}^{(1)}/\mathcal{U}$ 
for $\gamma_s = 1$ and $\gamma_a = 0$.
Therefore, similar sign inversions in the lateral force also arise in the classical context.

\section{Behavior of $U^{(1)}_{\text{vdW}}/{\cal{U}}$ for $0 < \theta < \pi/2$}
\label{app:orientacao}

\begin{figure}[h]
\centering  
\subfigure[]{\label{fig:gaussiana-orientacao-85}\epsfig{file=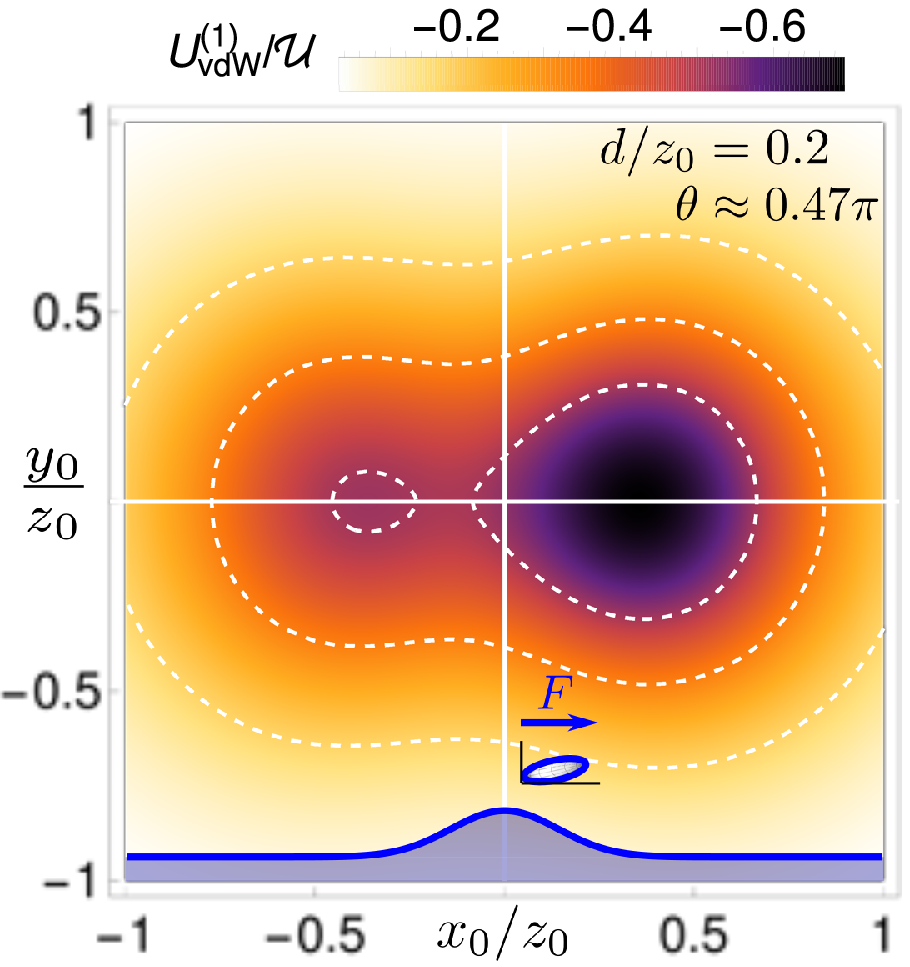, width=0.49 \linewidth}}
\hspace{0mm}
\subfigure[]{\label{fig:gaussiana-orientacao-60}\epsfig{file=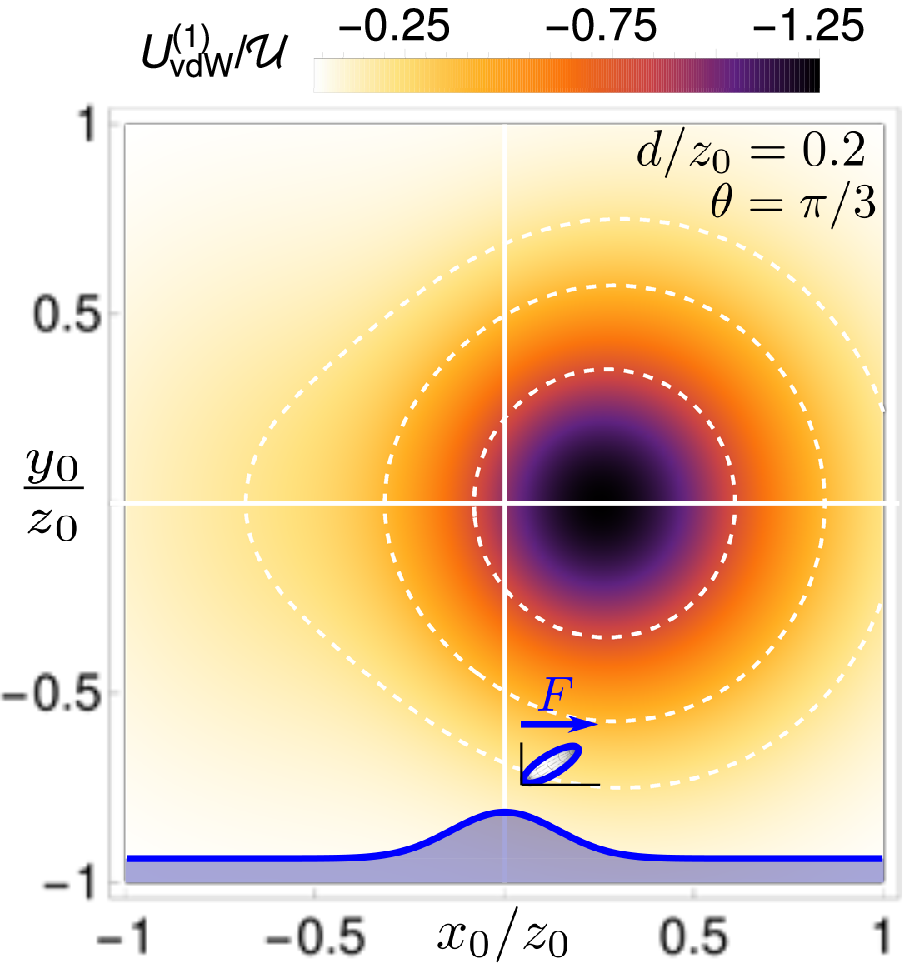, width=0.49 \linewidth}}
\caption{
Behavior of $U_{\mathrm{vdW}}^{(1)}/{\cal{U}}$ versus ${x_0}/{z_0}$ and ${y_0}/{z_0}$, for a particle fixed at $z = z_0$.
In each figure, we consider this particle characterized by $\gamma_{a}=0$ and $\gamma_{s}=0.6$, oriented such that $\phi=0$ and $\psi=0$, and the ratio ${d}/{z_0}=0.2$.
The insets in the bottom of the graphics illustrate the Gaussian profile projected on the plane $y=0$, a particle located at $(x_0,0,z_0)$, its orientation and the direction of the lateral vdW force acting on it.
From (a) to (b), there is a change on the particle orientation from $\theta=0.47\pi$ to $\theta=\pi/3$.
}
\label{fig:gaussiana}
\end{figure}

The sign inversion in the lateral vdW force is affected by the particle orientation.
Let us consider, for instance, a particle characterized by $\gamma_a = 0$ and $\gamma_s = 0.6$, oriented with $\phi=0,\psi=0$, and the ratio ${d}/{z_0}=0.2$.
When $\theta=\pi/2$ ($\hat{\mathbf{e}}_3^{\prime} = \hat{\mathbf{x}}$), we have the case shown in Fig. \ref{fig:gaussiana-x-02}, whereas, when $\theta=0$ $(\hat{\mathbf{e}}_3^{\prime} = \hat{\mathbf{z}})$, we have the case shown in Fig. \ref{fig:gaussiana-orientacao-0}.
In these two cases, $U^{(1)}_{\text{vdW}}/{\cal{U}}$ has a symmetric behavior.
On the other hand, when considering $0 < \theta < \pi/2$ one can see that such symmetry does not occur anymore.
In comparison to the case shown in Fig. \ref{fig:gaussiana-x-02}, if we make a small change in the particle orientation, for example, from $\theta=\pi/2$ to $\theta=0.47\pi$, one can see that the energy still has two minimum points, but one of them is, now, only a local minimum point [see Fig. \ref{fig:gaussiana-orientacao-85}].
However, if $\theta$ decreases further, for example, from $\theta=0.47\pi$ to $\theta=\pi/3$, we have only a global minimum point, which does not coincide with the peak of the Gaussian [see Fig. \ref{fig:gaussiana-orientacao-60}].

\section{Peak and valley regimes}
\label{app:regimes}

\begin{figure}[h]
\centering
\epsfig{file=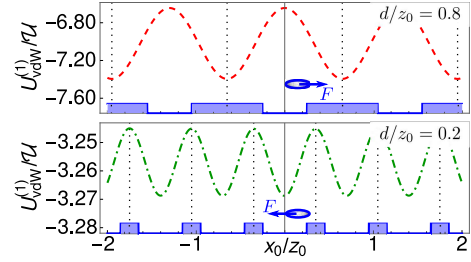, width=.7 \linewidth}
\caption{
Behavior of the ratio $U_{\mathrm{vdW}}^{(1)}/{\cal{U}}$ versus ${x_0}/{z_0}$, for a particle fixed at $z = z_0$, with $\gamma_{a}=0$, $\gamma_{s}=0.2$ and oriented with $\phi=0,\theta=\pi/2,\psi=0$.
We consider a periodic structure formed by several rectangular strips of width $d$ and distant $L$ from each other, so that $L/z_0 = 0.5$ (the surface profile is represented by the solid lines).
From top to bottom figure, we change the ratio ${d}/{z_0}$ from ${d}/{z_0}=0.8$ (dashed line) to ${d}/{z_0}=0.2$ (dot-dashed line).
In the top figure, note that the minimum points of the energy are over the centers of the protuberances, with such behavior called in Ref. \cite{Nogueira-PRA-2021} as peak regime.
On the other hand, in the bottom figure the minimum points are over the middle points between a pair of strips, with such behavior called in Ref. \cite{Nogueira-PRA-2021} as valley regime.
}\label{fig:regime-pico-vale}
\end{figure}

The result in Eq. \eqref{eq:kappa-strip} can be readily generalized for the case of $N$ strips of width $d$, distant by $L$ from each other.
In this case, the profile function is given by
\begin{align}
h(\mathbf{r}_{||}) & =a\sum_{k=1}^{N}\left\{ \Theta\left[x+\left(\frac{N+1}{2}-k\right)(L+d)+\frac{d}{2}\right]\right. \nonumber \\
 & \left.-\Theta\left[x+\left(\frac{N+1}{2}-k\right)(L+d)-\frac{d}{2}\right]\right\}, \label{eq:n-faixas}
\end{align}
so that, one can write 
\begin{align}
\mathcal{K}_{ij}^{(N)} & \left(\frac{x_{0}}{z_{0}},\frac{d}{z_{0}},\frac{L}{z_{0}}\right) \nonumber \\
 & =\sum_{k=1}^{N}\mathcal{K}_{ij}\left[\frac{x_{0}}{z_{0}}+\left(\frac{N+1}{2}-k\right)\left(\frac{L}{z_{0}}+\frac{d}{z_{0}}\right),\frac{d}{z_{0}}\right], \label{eq:kappa-nstrips}
\end{align}
with $\mathcal{K}_{ij}$ given by Eq. \eqref{eq:kappa-strip}. 
As $N$ increases, so that the strips tend to form a periodic structure, one can see these collective effects manifesting themselves as the peak and valley regimes (see Fig. \ref{fig:regime-pico-vale}), which were recently described in the literature \cite{Nogueira-PRA-2021,Queiroz-PRA-2021}.

%\bibliography{refs-casimir-polder}
%

\end{document}